\documentclass[%
article,
twocolumn,
nofootinbib,
amsmath,amssymb,
aps,
prl,
]{revtex4-2}
\usepackage[utf8]{inputenc}
\usepackage[english]{babel}
\usepackage{amsmath, amsfonts, amssymb}
\usepackage{nccmath}
\usepackage{mathrsfs}
\usepackage[mathscr]{euscript}
\usepackage{color}
\usepackage{amsthm}
\usepackage{psfrag}
\usepackage{graphicx}
\usepackage{url}
\usepackage[hyperfootnotes=false]{hyperref}
\usepackage{multirow}
\usepackage{makecell}
\usepackage{bbm}
\usepackage{graphicx}
\usepackage{dcolumn}
\usepackage{bm}

\def\d{\partial}

\renewcommand{\b}{\beta}

\newcommand{\e}{{\rm e}}

\renewcommand{\a}{\alpha}
\renewcommand{\b}{\beta}

\begin{document}
	
	
	\title{Effective Action for Dissipative and Nonholonomic
          Systems}
	
	\author{Afshin Besharat}%
	\author{Jury Radkovski}
        \author{Sergey Sibiryakov}
	\affiliation{
		Department of Physics and Astronomy, McMaster
  University, 1280 Main Street West, Hamilton, ON L8S 4M1, Canada\\
  Perimeter Institute for Theoretical Physics, Waterloo,
  Ontario, N2L 2Y5, Canada
	}

	\date{\today}
	
	\begin{abstract}
We show that the action of a dynamical system can be supplemented by
an effective action for its environment to reproduce arbitrary
coordinate dependent ohmic
dissipation and gyroscopic forces. The action is a generalization of the harmonic bath model and
describes a set of
massless interacting scalar fields in an auxiliary space
coupled to the original system at the boundary. 
A certain limit of the model implements nonholonomic
constraints. In the case of dynamics with nonlinearly realized
symmetries the effective action takes the form of a two-dimensional
nonlinear sigma-model. It provides a basis for
application of path integral methods to
general dissipative and nonholonomic systems. 
	\end{abstract}
	
	
\maketitle
	

\textbf{Introduction} -
Dissipation is ubiquitous in
nature. The standard way to account for it in the classical theory of
dynamical systems is by adding non-conservative forces $F_i$ to the
Euler--Lagrange equations of motion,
\begin{equation}
\label{eq:eomdis}
    \frac{d}{dt} \frac{\partial L}{\partial \dot{q}^{i}} -
    \frac{\partial L}{\partial q^{i}} = F_i \, , 
\end{equation}
where $L(q,\dot q)$ is the system Lagrangian, and $q^i$, $\dot q^i$ are
the generalized coordinates and velocities, respectively. An
important type of dissipation is ohmic dissipation when the extra
forces are linear in velocities,
\begin{equation}
\label{eq:ohmic}
    F_i = -\Gamma_{ij}(q) \,\dot q^j\equiv -\frac{\d F}{\d \dot q^i} \ .
\end{equation}
The dissipative coefficients $\Gamma_{ij}(q)$ form a positive-definite
symmetric matrix and are in general coordinate dependent. In the last
equality we have conventionally written the force as the derivative of
the Rayleigh function $F(q,\dot q)=(1/2) \Gamma_{ij}(q)\,\dot q^i\dot
q^j$. 

Fundamentally, the existence of dissipation is due to the interaction
of the system 
(referred to as {\it central system} below) 
with its environment, also called {\it
  reservoir} or {\it bath}. 
In many applications the microscopic nature of the
reservoir is not important and it can be
modeled as a collection of infinitely many harmonic oscillators
\cite{weiss2012quantum}. The action of the harmonic bath coupled to
the central system then provides an effective action, from which
Eq.~(\ref{eq:eomdis}) can be derived by means of the variational
principle. Yet more importantly, the effective action is key for the
application of the path integral methods used to study 
intrinsically quantum phenomena, such as tunneling
\cite{caldeira1981influence,caldeira1983quantum}, and other aspects of open systems in and
out of thermal equilibrium 
\cite{sieberer2016keldysh,de2017dynamics,kamenev2023field}.

However, as we discuss below, the harmonic bath model fails
in the case when dissipation coefficients $\Gamma_{ij}(q)$ have general
dependence on the system
coordinates.
The
purpose of this {\it Letter} is to provide a reservoir model for
this case. As a byproduct we also obtain the description of arbitrary
gyroscopic forces. Before describing the model, let us
discuss two broad classes of situations where the dependence of
$\Gamma_{ij}$ on $q$ is
essential. 

\textbf{Dynamics on cosets} - The first class are systems whose
configuration space represents a group manifold or, more generally, a
coset space, and whose dynamics enjoy non-linearly realized
symmetries. Many physically relevant systems can
be cast in this form, from dynamics of a rigid body, to
hydrodynamics \cite{arnold2008topological,marsden2013introduction}.
They appear in particle physics and condensed matter as a consequence of spontaneous symmetry breaking
\cite{Burgess:1998ku,Brauner:2010wm}.  
Development of an effective action for such systems in dissipative
environment, besides conceptual interest, is motivated by numerous
potential applications, for example to Brownian motion 
of stiff 
polymers \cite{li2004diffusion}, as well as micro and nanoparticles of
various shapes  
\cite{duggal2006dynamics,han2006brownian,kraft2013brownian,
zhang2019diffusion}.    

Following the standard coset construction \cite{callan1969structure,Burgess:1998ku,Brauner:2010wm}, 
we consider a Lie
group $\mathbb{G}$ and its subgroup $\mathbb{H}$. The generators of
$\mathbb{G}$ are chosen in such a way that the first $A$ of them span
the algebra of $\mathbb{H}$, we denote them by $H_a$,
$1\leq a\leq A$. The rest of the $\mathbb{G}$-generators are called
{\it broken} and will be denoted by $\hat G_i$,  $1\leq i\leq I$. 
The coset $\mathbb{G}/\mathbb{H}$ representing the configuration space
of the system can be identified with the group elements of the
form\footnote{We assume summation over repeated indices.}
\begin{equation}
\label{eq:grepr}
    \hat g(q)=\e^{q^i \hat G_i} \ .
\end{equation}
The action of a group element $g$ 
on the coordinates $q\mapsto \tilde q$ is given by
the right multiplication,
\begin{equation}
\label{eq:groupact}
g\cdot\hat g(q)=\hat g(\tilde q)\cdot h~,~~~~~g\in \mathbb{G},~h\in
\mathbb{H}\;. 
\end{equation}
Next, one constructs the Cartan form,
\begin{equation}
\label{eq:GCartan}
   \hat g^{-1}d\hat g =\Omega^i_j(q) dq^j\,\hat G_i +
\Omega^a_j(q) dq^j\, H_a\ \;,
\end{equation}
and extracts from it the {\it covariant velocities} 
\begin{equation}
\label{eq:covD2d}
    D_tq^i=\Omega^i_j(q)\,\dot q^j \ .
\end{equation}
Unlike the ordinary velocities $\dot q^i$, the covariant
velocities 
transform linearly under 
(\ref{eq:groupact}). They form a linear representation of the
subgroup $\mathbb{H}$. 

If the dynamics of the system are to respect the
symmetry (\ref{eq:groupact}), its Lagrangian
and Rayleigh function must 
be invariants
constructed from the covariant velocities.\footnote{Up to
  possible Wess--Zumino--Witten terms 
\cite{wess1971consequences,witten1983global,Goon:2012dy}.} 
Thus we have,
\begin{equation}
F=\tfrac{1}{2}\gamma_{ij} D_tq^i D_t q^j=\tfrac{1}{2}\gamma_{ij}
\Omega_k^i(q)\Omega_l^j(q)\,\dot q^k\dot q^l\;, 
\end{equation}   
where $\gamma_{ij}$ is a constant invariant tensor in the relevant
representation of  
$\mathbb{H}$. In the simplest case when $\mathbb{H}$ is empty 
($\mathbb{G}$ is fully broken) $\gamma_{ij}$ 
is arbitrary, provided it is symmetric and positive. For a general
non-Abelian coset the coefficients of the Cartan form
satisfy
\begin{equation}
\label{eq:Omnonint}
\frac{\d \Omega^k_i}{\d q^j}-\frac{\d \Omega^k_j}{\d q^i}\neq 0\;,
\end{equation}
so their dependence on coordinates cannot be avoided by any choice of
variables,
implying the coordinate dependence of
the dissipative coefficients 
$\Gamma_{kl}=\gamma_{ij}\Omega_k^i(q)\Omega_l^j(q)$. 

\textbf{Nonholonomic systems} - The second class are systems with
constraints on coordinates and velocities,\footnote{We only consider constraints linear in velocities.}
\begin{equation}
\label{eq:nhconstr}
    c^{\alpha}_{i}(q)\, \dot{q}^{i} = 0 \, , \quad \alpha = 1,\dots n <I \ ,
\end{equation}
such that they cannot be integrated into constraints only on
coordinates. In other words, Eq.~(\ref{eq:nhconstr}) is not equivalent
to a set of constraints of the form $\dot\varphi^\alpha(q)=0$. Clearly,
this requires 
\begin{equation}
\label{eq:cnonint}
\frac{\d c^\alpha_i}{\d q^j}-\frac{\d c^\alpha_j}{\d q^i}\neq 0\;.
\end{equation}
These systems are called nonholonomic and typical examples
include rolling of a disk or a ball on a hard surface.
Their classical
dynamics is well developed and is summarized
in excellent textbooks, e.g. \cite{ne_mark2004dynamics,arnold2006mathematical}. 
Quantization, however, remains an open problem. It was addressed in 
\cite{bloch2008quantization,fernandez2018quantum,fernandez2022quantizing} and presents a 
growing interest due to development of molecular machines 
\cite{shirai2005directional,grill2007rolling,erbas2015artificial}. 

Typically, the equations of motion for nonholonomic systems are
derived from a modified variational principle restricted to admissible
variations $\delta q^i$ satisfying the constraints $c^\alpha_i\,\delta
q^i=0$. This leads to the appearance of {\it reaction forces} on the
r.h.s. of the Euler--Lagrange equations (\ref{eq:eomdis}),
\begin{equation}
\label{eq:Leqgen}
   F_i = \lambda_{\alpha} c^{\alpha}_{i}(q) \ ,
\end{equation} 
where $\lambda_\alpha(t)$ are Lagrange multipliers.\footnote{Note that
adding the constraints (\ref{eq:nhconstr}) with Lagrange multipliers
into the Lagrangian, instead of the equations of motion, would not
reproduce the correct nonholonomic dynamics. Instead, one would obtain
a so-called vakonomic system \cite{arnold2006mathematical}.} 
Due to the constraints (\ref{eq:nhconstr}), the reaction forces
do not produce any work, $F_i\dot q^i=0$, so 
nonholonomic systems are not truly dissipative. However, they are
closely related through the following construction
\cite{ne_mark2004dynamics,arnold2006mathematical}. Consider a
dissipative system with the Rayleigh function  
\begin{equation}
\label{eq:Fnh}
    F = \tfrac{1}{2}\gamma\, c^{\alpha}_{i}(q) 
c^{\alpha}_{j}(q) \,\dot q^i\dot q^j
\end{equation}
and take the limit $\gamma\to+\infty$. 
The friction associated with the linear combinations of velocities 
$c^{\alpha}_{j} \dot{q}^{j}$ becomes very strong and the corresponding
combinations quickly die out rendering the constraints
(\ref{eq:nhconstr}). On the other hand,
the products $\gamma c^{\alpha}_{i} \dot{q}^{i}$
remain finite and become independent variables --- the Lagrange
multipliers of Eq.~(\ref{eq:Leqgen}). Thus,
the nonholonomic dynamics
can be viewed as the limit 
of infinitely strong viscous friction along
the constrained directions.

\textbf{The reservoir model} - Our starting point is the model used
in \cite{unruh1989reduction} to study environment-induced 
decoherence. It represents the reservoir as a free massless scalar
field $\xi(t,z)$ in one-dimensional space ({\it bulk}) coupled to the
central system at a single point $z=0$ ({\it boundary}), and is
equivalent to the more common independent-oscillator
model \cite{caldeira1983quantum}. Its straightforward generalization
for a central system with several degrees of freedom
requires equal number of fields and leads to the following action, 
\begin{equation}
    \label{eq:SCL}
	S =\int\limits_{z=0} dt \, \big(L(q,\dot q)-\beta_i^j  q^i \dot
\xi_j\big) 
 + \int\limits_{z>0} dt dz \, \frac{1}{2} \d_\mu \xi_i\d^\mu\xi_i \;,
\end{equation}
where $\beta_i^j$ are constant couplings; in the last term we sum
over indices $\mu=t,z$ with the Lorentzian metric
$\eta^{\mu\nu}={\rm diag}(1,-1)$. 
Importantly, the coordinate $z$ here is not a 
physical dimension, but is introduced merely to
parameterize the internal dissipative degrees of freedom.
By taking variation, one derives
the dissipative forces, as well as the equations for the fields,
\begin{equation}
\label{eq:CLeqs}
F_i=-\beta_i^j\dot \xi_j\big|~,~~~~
\d_\mu\d^\mu \xi_i=0~,~~~~\d_z\xi_i\big|=-\beta^i_j\dot q^j\;,
\end{equation}
where the vertical bar means fields evaluated at $z=0$.
The dissipative dynamics is obtained by imposing
outgoing boundary conditions on the bulk fields  
which singles out the solutions of the form
$\xi_i(t,z)=\bar\xi_i(t-z)$. This implies $\d_z\xi_i=-\d_t\xi_i$ and
combining the first and third equations in (\ref{eq:CLeqs}) we obtain
the forces (\ref{eq:ohmic}) with
$\Gamma_{ij}=\beta_i^k\beta_j^k$. 
Note that coupling $q^i$ to $\dot\xi_i$, rather than the fields themselves, is essential for getting the response local in time.

The above construction fails for general coordinate dependent
dissipation. As long as we want to preserve the harmonic nature of the
bath, the only option is to generalize its coupling to the
central system, 
$\beta_i^j q^i\dot\xi_j\mapsto
\beta^j(q)\dot\xi_j$ with some arbitrary functions
$\beta^i(q)$. Repeating the above derivation we then obtain the
dissipative coefficients 
$\Gamma_{ij}=(\d\beta^k/\d q^i)(\d\beta^k/\d q^j)$ which, however, do
not have the form needed for coset or nonholonomic systems due to the
non-integrability properties (\ref{eq:Omnonint}), (\ref{eq:cnonint}).

This failure can be also understood from the symmetry perspective. The
system-reservoir coupling in (\ref{eq:SCL}) 
changes by a total time-derivative under the shifts of the
coordinates 
$q^i(t)\mapsto q^i(t)+a^i$.
This property ensures that the dissipative force is invariant
under the coordinate shifts, as it should be for the case of constant
$\Gamma_{ij}$. In the case of a general non-Abelian coset, however, we do not
have at our disposal any functions $\beta^i(q)$ 
invariant or changing by a constant under the group transformations and
hence we cannot construct any system-reservoir coupling that would
preserve the symmetry of the problem.\footnote{Integrating by parts the
interaction term in (\ref{eq:SCL}) and replacing $\dot q^i$ with
the covariant derivative $D_tq^i$ does not help. We then have $\xi_i$, instead of $\dot \xi_i$ in the coupling, 
which leads to forces
$F_i$ with non-local memory of the past motion of the system.}

To resolve the issue, we apply a duality transformation to
the action (\ref{eq:SCL}). Performing a change of variables
$\tilde\xi_i=\beta_i^j\xi_j$ and integrating in a set of vectors
$\chi^{\mu\,i}$, it can be rewritten as  
\begin{align}
    \label{eq:SCLeqv3}
        S\! =\!\!\!\int\limits_{z=0}\!\!\! dt\,\big( L(q,\dot q)-q^i\dot{\tilde\xi}_i\big)
+\!\!\!\int\limits_{z>0}\!\!\! dt dz \, \bigg(\!\chi^{\mu\,i} \d_\mu\tilde\xi_i
	-\frac{\Gamma_{ij}}{2}\chi^{\mu\,i}\chi_\mu^j\!\bigg) .
\end{align}
We now integrate out $\tilde \xi_i$, which gives two equations,
\begin{equation}
\label{eq:xiconstr}
\quad \d_\mu \chi^{\mu\, i}=0\;,~~~~  \chi^{z\,i}\big|=\dot q^i\;.
\end{equation}
The first one implies that $\chi^{\mu\,i}$ are expressed through
gradients of scalar functions,
\begin{equation}
\label{eq:chisol}
    \chi^{\mu\,i}=-\epsilon^{\mu\nu}\d_\nu\chi^i \ ,
\end{equation}
where $\epsilon^{\mu\nu}$ is the two-dimensional Levi--Civita symbol,
$\epsilon^{tz}=1$.
The second equation then reduces to $\dot\chi^i\big|=\dot q^i$. 
Using the fact that the fields
$\chi^i$ are defined up to a constant,
we can remove any offset between them and 
$q^i$ on the boundary, and obtain  
\begin{equation}
\label{eq:chibc}
\chi^i\big|=q^i \ .
\end{equation}
Substituting (\ref{eq:chisol}) back into (\ref{eq:SCLeqv3}) we arrive
at the action 
\begin{equation}
\label{eq:S2dnew}
    S=\int dt\, L(q,\dot q)+\int_{z>0} dt
    dz\,\frac{1}{2}\Gamma_{ij}\d_\mu \chi^i\d^\mu \chi^j 
\end{equation} 
with the boundary conditions (\ref{eq:chibc}). 
For a single degree of freedom $q$ this action first
appeared in \cite{lamb1900peculiarity} and was used in
\cite{ford1988quantum} 
for the
derivation of the  
quantum Langevin equation. More recently, it was extended to describe
linear response in dissipative media \cite{figotin2007hamiltonian}. 

So far, we have assumed the dissipative coefficients $\Gamma_{ij}$
to be constant. However, the action (\ref{eq:S2dnew}) 
admits a natural generalization.
Relation (\ref{eq:chibc}) suggests thinking of the fields $\chi^i$ as
extensions of the original system coordinates into the bulk. Then, to
describe coordinate dependent dissipation, we
simply need to promote the coefficients in (\ref{eq:S2dnew}) to the
functions of $\chi$,
\begin{equation}
\Gamma_{ij}\mapsto \Gamma_{ij}(\chi)\;.
\end{equation}
Note that this makes the effective reservoir fields
self-interacting. It is a necessary price to pay for modeling
coordinate dependent friction.

This is not yet the whole story. We can add to the reservoir action a
time-reversal breaking term
\begin{equation}
\int_{z>0} dt dz\, \frac{1}{2}\Upsilon_{ij}(\chi)
\epsilon^{\mu \nu} \partial_{\mu} \chi^{i} \partial_{\nu} \chi^{j}
\end{equation}
with antisymmetric coefficients $\Upsilon_{ij}(\chi)$. If
$\Upsilon_{ij}$ are constant, this term is a total derivative and reduces to
the boundary term $\int dt\, \Upsilon_{ij}q^i\dot q^j$ of the
Wess--Zumino--Witten type
\cite{wess1971consequences,witten1983global,Goon:2012dy}. 
However, for the general field dependent coefficients such reduction
is impossible. 

Combining all the above ingredients, we write down the action of our
reservoir model:
\begin{align}
\label{eq:themodel}
S=\int dt\,L(q,\dot q)
+\int_{z>0}\!\!dt&dz\,\frac{1}{2}\Big(\Gamma_{ij}(\chi)\d_\mu\chi^i\d^\mu\chi^j
\nonumber\\
&+\Upsilon_{ij}(\chi)\epsilon^{\mu\nu}\d_\mu\chi^i\d_\nu\chi^j\Big)\,.
\end{align}
Let us verify that it reproduces the desired equations. Taking its
variation and accounting for the relation (\ref{eq:chibc}) we obtain
in the bulk and on the boundary:
\begin{align}
\label{eq:newLeq}
&\d_\mu\Big(\Gamma_{ij}\d^\mu\chi^j+\Upsilon_{ij}\epsilon^{\mu\nu}\d_\nu\chi^j\Big)
\nonumber\\ 
&-\frac{1}{2}\frac{\d \Gamma_{jk}}{\d \chi^i}
\d_\mu\chi^j\d^\mu\chi^k
-\frac{1}{2}\frac{\d \Upsilon_{jk}}{\d \chi^i}
\epsilon^{\mu \nu} \d_\mu \chi^j \d_\nu \chi^k=0\;, \\
\label{eq:newbc}
&F_i=\big(\Gamma_{ij}\d_z \chi^j+\Upsilon_{ij}\dot
\chi^j\big)\big|\;. 
\end{align}
Though the bulk equation (\ref{eq:newLeq}) looks complicated, it still
admits purely outgoing solutions $\chi^i(t,z)=\bar\chi^i(t-z)$ with
arbitrary functions $\bar\chi^i$. The conditions (\ref{eq:chibc}) then
fix $\bar\chi^i(t)=q^i(t)$ and Eq.~(\ref{eq:newbc}) reduces to 
\begin{equation}
\label{eq:correq}
  F_i=-\Gamma_{ij}(q)\,\dot q^j+\Upsilon_{ij}(q)\,\dot q^j \ .
\end{equation}
The first term gives the sought-after dissipative forces
(\ref{eq:ohmic}), whereas the second term describes arbitrary
gyroscopic forces that arise if the environment breaks
time-reversal symmetry, e.g. by magnetization or rotation. The effective reservoir action (\ref{eq:themodel}) represents our main result. To the best of our knowledge, this action never appeared in the literature before. It covers a much broader class of systems than the original harmonic bath (\ref{eq:SCL}).

\textbf{Nonholonomic limit} - To describe a nonholonomic system, we
replace 
$\Gamma_{ i j} \mapsto \Gamma_{i j} + \gamma c^{\a}_{i} c^{\a}_{j}$
and send $\gamma$ to infinity. 
The resulting action can be obtained in a closed form by integrating
in a set of auxiliary vectors $\lambda^\mu_\a(t,z)$. Omitting for
simplicity the time-reversal breaking term, we write:
\begin{align}
    \label{eq:ScosetNH1}
        S_{\rm nh} =& \int dt \, L(q,\dot q) 
+ \int_{z>0}dtdz\,\bigg(\frac{1}{2} \Gamma_{ i j}(\chi)
        \partial_{\mu} \chi^{i} \partial^{\mu} \chi^{j} \nonumber \\
        &- \lambda^\mu_\a c^{\alpha}_{i}(\chi) \d_\mu \chi^{i}
        -\frac{1}{2\gamma} \lambda_\a^\mu\lambda_{\mu\,\b}\bigg) \ . 
\end{align}
In the limit $\gamma\to\infty$ the last term disappears and the fields
$\lambda_\a^\mu$ become Lagrange multipliers enforcing the constraints 
$c_i^\a(\chi)\d_\mu\chi^i=0$. Since $\chi^i$ coincide with $q^i$ on
the boundary, this implies the nonholonomic constraints
(\ref{eq:nhconstr}). The remaining equations also come out
right. Varying (\ref{eq:ScosetNH1}) with respect to $\chi^i$ and
substituting the outgoing solution into the boundary equation, we obtain
the force, 
\begin{equation}
\label{LeqNH1}
    F_i= -\Gamma_{ij}(q)\, \dot q^j+c^{\a}_{i}(q)\, \lambda_{\a}^{z}\big|\;.
\end{equation} 
The last term gives precisely the reaction forces along the constrained
directions (\ref{eq:Leqgen}), with $\lambda_\a^z\big|$ playing the
role of the Lagrange multipliers from the standard approach. The first
term describes friction along the unconstrained directions. Note that in our 
approach it cannot be set to zero without making
the bulk action degenerate.

\textbf{Example} - The model takes a particularly simple form for
motion on cosets: 
\begin{align}
\label{eq:Scoset}
S_{\rm coset}&=\int dt\, L(D_t q)
\nonumber\\
&+\int_{z>0}\! dtdz\,\frac{1}{2} \bigg(\gamma_{ij} \eta^{\mu \nu} +
\upsilon_{ij}\epsilon^{\mu \nu}\bigg) D_\mu \chi^i D_\nu \chi^j \,, 
\end{align}
where $D_\mu\chi^i\equiv \Omega_j^i(\chi)\d_\mu\chi^j$ are covariant
derivatives of the fields $\chi^i$, and $\gamma_{ij}$, $\upsilon_{ij}$
are constants. One recognizes in the bulk term the action of a
two-dimensional nonlinear sigma-model \cite{zinn2021quantum}. 
It is
the most general local action that can be written using the first
derivatives of the fields $\chi^i$ and invariant under
the group $\mathbb{G}$.

\begin{figure}[tb]
\begin{center}
\includegraphics[scale=0.45]{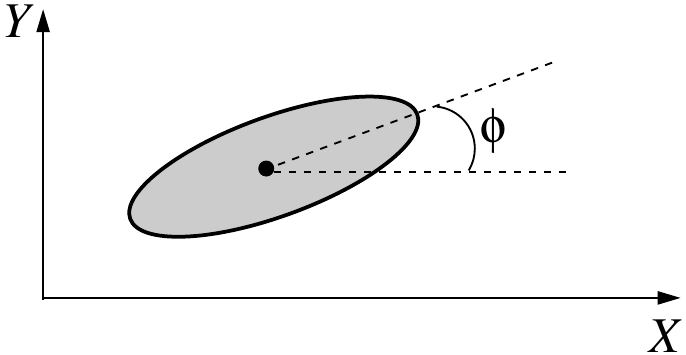}
\end{center}
\caption{\label{fig:sleigh} Oblong particle on a plane.}
\end{figure}

Let us illustrate this construction in the case of an oblong
particle moving on a two-dimensional plane
in a viscous medium
\cite{li2004diffusion,duggal2006dynamics,han2006brownian}. Its
position is described by the center-of-mass coordinates
$X$, $Y$ and the orientation angle $\phi$, see Fig.~\ref{fig:sleigh}. The friction
coefficients are different in the directions along and perpendicular
to the particle's main axis. The configuration space coincides with the
group of isometries of the Euclidean plane $ISO(2)$ which has
two generators of translations $P_X$,
$P_Y$ and a rotation generator $J$. The commutation relations are:
\begin{equation}
\label{eq:ISOcomm}
    [P_X,P_Y]=0~,~~~~[P_X,J]=-P_Y~,~~~~[P_Y,J]=P_X\;.
\end{equation}
All generators are broken. We parameterize the group elements 
as\footnote{This parameterization slightly differs from
  Eq.~(\ref{eq:grepr}) and makes the calculations simpler.}
\begin{equation}
\label{eqISOel}
    g(X,Y,\phi)=\e^{XP_X+YP_Y}\e^{\phi J}\;.
\end{equation}
From the Cartan form we get the 
covariant derivatives: 
\begin{align}
&D_t X=\dot X\cos\phi +\dot Y\sin \phi~,~~\\
&D_t Y=-\dot X\sin\phi +\dot Y\cos \phi~,~~D_t\phi=\dot\phi\;.
\end{align}
The Lagrangian coincides with the kinetic energy of the particle and
is $ISO(2)$ invariant,
\begin{equation}
    L \!=\! \frac{m }{2}(\dot{X}^2 + \dot{Y}^2) + \frac{{\cal I} }{2}\dot{\phi}^2
\!\!=\! \frac{m}{2}\big[ (D_{t}X)^2+ (D_{t} Y)^2\big]\!+
\frac{\cal I}{2} (D_t\phi)^2,
\end{equation}
where $m$, ${\cal I}$ are the particle mass and moment of
inertia. 

If the viscous medium is homogeneous and
isotropic, the effective reservoir action must also enjoy $ISO(2)$
symmetry. To implement it, we introduce the fields $\Xi(t,z)$,
$\Psi(t,z)$, and $\Phi(t,z)$, such that at $z=0$ they coincide with
$X(t)$, $Y(t)$ and $\phi(t)$, respectively. We recall that
the coordinate $z$ is not a physical dimension. Rather, it
parameterizes the internal degrees of freedom of the particle and
medium responsible for dissipation. The effective bath action then reads,
\begin{align}
S_{\rm bath}=\int_{z>0}dtdz\,
\frac{1}{2}\big(&\gamma_\parallel D_\mu\Xi D^\mu\Xi
+\gamma_\perp D_\mu\Psi D^\mu\Psi
\nonumber\\
&+\gamma_\phi D_\mu\Phi D^\mu\Phi\big)\;,
\end{align}
where
\begin{align}
&D_\mu \Xi=\cos\Phi\,\d_\mu\Xi +\sin \Phi\,\d_\mu \Psi~,~~\\
&D_\mu\Psi=-\sin\Phi\,\d_\mu\Xi +\cos \Phi\,\d_\mu\Psi 
~,~~D_\mu\Phi=\d_\mu\Phi\;.
\end{align}
We observe that even in this relatively simple case the bath action is
nonlinear if $\gamma_\perp\neq \gamma_\parallel$. 
In the limit $\gamma_\perp\to+\infty$ we obtain a
particle that is constrained to move along its major axis. This is the 
simplest 
nonholonomic system known as {\it Chaplygin sleigh}.


\textbf{Discussion} - We have presented a reservoir model
(\ref{eq:themodel}) for systems with general coordinate dependent
ohmic dissipation and gyroscopic forces. In geometric terms, it can
be viewed as a semi-infinite string moving on a curved
target space. The coordinate along the string labels the continuum of
internal reservoir degrees of freedom. In a  certain limit, the model
reproduces nonholonomic constraints. If dynamics obey
nonlinearly realized symmetries, the reservoir takes the form of a
two-dimensional nonlinear sigma-model. 

The model has a vast range of potential applications. It provides a basis for development of path integral methods and quantization\footnote{Note in this connection that the
sigma-model in 2d is renormalizable.} in a broad class of dissipative and nonholonomic systems. One possible direction is derivation of classical and quantum Langevin equations for state-dependent diffusion \cite{Lau_2007} and Brownian motion of extended impurities \cite{li2004diffusion,duggal2006dynamics,han2006brownian,kraft2013brownian,zhang2019diffusion}, including systematic treatment of the multiplicative and non-Gaussian noise. 
Another interesting direction is generalization of the model to systems with an infinite number of degrees of freedom.
Here promising arenas for applications are 
dissipative hydrodynamics \cite{Liu:2018kfw} and open effective field theories \cite{Burgess:2022rdo}. 

An important question is to what extent the model (\ref{eq:themodel})
is universal beyond the classical equations. Does it capture the relevant
properties of any ohmic
environment? The unique structure of the model for coset dynamics
suggests that, at least in this case, it is indeed
universal in the sense of effective theory.
Namely, we conjecture that the
correlators of the $q^i$-variables obtained from the action (\ref{eq:Scoset}) reproduce the most general long-time behavior of correlators in  
dissipative systems with given non-linear symmetries. 
We leave the exploration of this conjecture for the future. An
interesting related work is construction of the effective
Schwinger--Keldysh functional for cosets \cite{Akyuz:2023lsm}.

One of the assumptions of our model is locality in the auxiliary
dimension $z$. This property is reminiscent of holography \cite{Jana:2020vyx} and it is tempting
to speculate that it arises as a consequence of the large number of
bath degrees of freedom \cite{Sundrum:2011ic}. Let us note that
we have not discussed the metric in the internal $(t,z)$ spacetime. At
the classical level, it drops out due to the classical Weyl
invariance of the action (\ref{eq:themodel}). In quantum theory,
however, it may become relevant due to the conformal anomaly. 


\textbf{Acknowledgments} -
We are indebted to Cliff Burgess, Cecile Fradin,
Ted Jacobson, Andrew Kovachik, 
Sung-Sik Lee, Duncan O'Dell,  
and Oriol Pujolas
for valuable discussions. 
The work is supported by
the 
Natural Sciences and Engineering Research Council (NSERC) of Canada.
Research at Perimeter Institute is supported in part by the Government
of Canada through the Department of Innovation, Science and Economic
Development Canada and by the Province of Ontario through the Ministry
of Colleges and Universities. 

\bibliography{refs.bib}

\end{document}